\documentclass[aps,twocolumn,prl,superscriptaddress,preprintnumbers,showpacs,tightenlines]{revtex4}
\usepackage{amssymb}
\usepackage{times}
\usepackage{graphicx}
\usepackage{epsfig}
\usepackage{color,soul}
\usepackage{multirow}
\usepackage{threeparttable}
\usepackage{bm}
\UseRawInputEncoding

\begin{document}
\title{Hydrogen diffusion in ceria: solid state NMR, combined scattering and spectroscopic studies, and {\it ab initio} calculations}
\author{Zhichao Zhu$^+$}
\affiliation{Institute of Nuclear Physics and Chemistry, China Academy of Engineering Physics (CAEP), Mianyang 621999, P R China}
\author{Xin Li$^+$}
\affiliation{Institute of Nuclear Physics and Chemistry, China Academy of Engineering Physics (CAEP), Mianyang 621999, P R China}
\author{Ziru Ma$^+$}
\affiliation{Institute of Nuclear Physics and Chemistry, China Academy of Engineering Physics (CAEP), Mianyang 621999, P R China}
\author{Yuanhua Xia}
\affiliation{Institute of Nuclear Physics and Chemistry, China Academy of Engineering Physics (CAEP), Mianyang 621999, P R China}
\author{Ruizhi Qiu}
\affiliation{Science and Technology on Surface Physics and Chemistry Laboratory, Mianyang 621907, P R China}
\author{Baijiang Lv}
\affiliation{Institute of Nuclear Physics and Chemistry, China Academy of Engineering Physics (CAEP), Mianyang 621999, P R China}
\author{Guanyun Yan}
\affiliation{Institute of Nuclear Physics and Chemistry, China Academy of Engineering Physics (CAEP), Mianyang 621999, P R China}
\author{Jianrong Zeng}
\affiliation{Shanghai Synchrotron Radiation Facility, Shanghai Advanced Research Institute, Chinese Academy of Sciences, Shanghai 201210, P R China}
\affiliation{Shanghai Institute of Applied Physics, Chinese Academy of Sciences, Shanghai 201800, P R China}
\author{Long Yang}
\affiliation{Interdisciplinary Materials Research Center, School of Materials Science and Engineering, Tongji University, Shanghai 201804, P R China}
\author{Jianbo Ma}
\affiliation{Shanghai Synchrotron Radiation Facility, Shanghai Advanced Research Institute, Chinese Academy of Sciences, Shanghai 201210, P R China}
\author{Benqiong Liu}
\email{liubenqiong@caep.cn}
\affiliation{Institute of Nuclear Physics and Chemistry, China Academy of Engineering Physics (CAEP), Mianyang 621999, P R China}
\author{Guangai Sun}
\email{guangaisun_80@163.com}
\affiliation{Institute of Nuclear Physics and Chemistry, China Academy of Engineering Physics (CAEP), Mianyang 621999, P R China}

$[^+]$ These authors contributed equally to this work.


\begin{abstract}
Ceria has been extensively studied since it has many applications in diverse research fields. However, the mechanism of the hydrogen dynamics, especially the diffusion kinetics on a microscopic level is still unclear as the experimental data has been very limited. In this work, the CeO$_2$-H interaction has been comprehensively studied by a combination of $^1$H NMR transverse relaxation time ($T_2$) measurement, neutron powder diffraction, quasi-elastic neutron scattering (QENS), X-ray total scattering, and {\it ab initio} calculations. Based on QENS measurements, the first direct evidence for hydrogen jump diffusions of the Chudley-Elliot type in the bulk ceria has been given, with a jump distance of $\sim$3.98 {\AA}. The theoretically calculated activation energy barriers $E_\mathrm{a}$ for hydrogen diffusion are relatively low ($<0.1$ eV), further supporting that such hopping can readily occur. A larger barrier value of $E_\mathrm{a}\sim$0.2 eV is directly estimated by $T_2$ NMR data, suggesting possible slower hydrogen dynamics with the pre-exponential factor $D_0$ of diffusion coefficient $\sim10^{-11}$ m$^2/$s.
\end{abstract}

\maketitle

\section{\uppercase\expandafter{\romannumeral1}. INTRODUCTION}

Ceria (CeO$_2$) has received considerable research interest due to its wide range of technological applications in catalysis \cite{T.Montini2016,N.A.M.Fadzil2018}, solid oxide fuel cells (SOFC) \cite{L.Vivier2010}, and biomedical area \cite{L.He2015}. It has also been expected to be a protective surface coating on alloys to hinder the hydrogen penetration into the bulk, protecting materials from hydrogen embrittlement \cite{D.Marrocchelli2012}. Many of the key properties of ceria that contribute to its success in catalysis are due to the exceptional capability of oxygen storage-and-release, as the cerium ions Ce$^{4+}$ and Ce$^{3+}$ can readily convert in the reduction/oxidation reaction. Therefore, the study of ceria in the presence of a reducing gas like H$_2$ is of significant importance to clarify the corresponding reduction mechanism.

The CeO$_2$-H$_2$ interaction is extremely complicated. Whether hydrogen atoms form hydroxyl (OH) or hydride species (H$_y$CeO$_2$, 0$\le y\le$1, {\it via} Ce-H functional group) on the surface as well as in the bulk has long been disputed. Li {\it et al.} \cite{Z.Li2019,Z.Li2021} demonstrated that on a stoichiometric ceria surface, H$_2$ only dissociated to form surface hydroxyls at elevated temperatures, while in reduced ceria CeO$_{2-x}$ both hydroxyls and hydrides formed on the surface as well as in the bulk. The oxygen vacancies played a key role in the formation of hydrides \cite{K.Werner2017,D.Schweke2020}, with the oxidation of Ce$^{3+}$ to Ce$^{4+}$ and one electron transferring to a hydrogen atom H$^-$. Matsukawa {\it et al.} \cite{T.Matsukawa2018,T.Matsukawa2021} investigated the redox reaction and determined the crystal structure of ceria at various temperatures under hydrogen by neutron diffraction, which revealed that ceria only transformed to the metastable cubic oxyhydroxide structure CeO$_2$H$_y$ (0$\le y\le$1, {\it via} OH formation in the lattice), while the cubic hydride CeH$_x$ could not be confirmed. For the hypothetically possible hydride ceria, no evidence could be found \cite{M.Gruenbacher2018} by powder X-ray diffraction, although ceria clearly incorporated hydrogen and two reduced phases were identified, {\it i.e.}, the weakly reduced phase with fluorite structure, and the strongly reduced phase with triclinic structure. In a recent study of ceria thin films \cite{W.Mao2024}, the large H content in the near-surface region was identified as hydroxyl, whereas stably bound hydrogen in the bulk of the films was nearly uniformly distributed and of very low concentration, thus the bulk H species might be strongly bound to defects in the films rather than the formation of hydrides.

In order to identify the hydrogen configuration in ceria, except for the direct structure determination by X-ray/neutron, there is another experimental approach that the corresponding vibrational modes of Ce-H, and different coordinated OH species can be assigned by vibrational spectroscopic techniques, such as infrared \cite{M.Gruenbacher2018,J.L.G.Fierro1985,A.Badri1996}, Raman, inelastic neutron scattering (INS) \cite{P.C.H.Mitchell2005,R.Juarez2010}, and normally combined with density functional theory (DFT) calculations. For examples, Wu {\it et al.} \cite{Z.Wu2017} found first direct spectroscopy evidence of the formation of Ce-H upon H$_2$-CeO$_2$ interaction by {\it in situ} INS technique. The vibrational modes of both OH and Ce-H functional groups were also detected by a recent INS measurement \cite{T.Matsukawa2021}. There were three intense INS peaks observed in the CeO$_2$-H sample, and the peak at $\sim$ 580 cm$^{-1}$ can be assigned to the OH group, and the other two peaks at $\sim$495 and $\sim$ 650 cm$^{-1}$ were attributed to the Ce-H species. As the hydride structure was not confirmed by neutron powder diffraction, they suggested that the Ce-H species were not created in the bulk ceria but existed on surface as a local structure, which probably could be revealed by pair distribution function (PDF) analysis.

Despite the fact that the CeO$_2$-H$_2$ interaction has been extensively investigated both experimentally \cite{J.L.G.Fierro1985,S.Bernal1993} and theoretically \cite{K.Sohlberg2001,M.Garcia-Melchor2014,D.Fernandez-Torre2014,F.R.Negreiros2015}, the microscopic mechanism for hydrogen transport in ceria has not been fully understood yet. Theoretically, Chafi {\it et al.} \cite{Z.Chafi2009} first reported that hydrogen atoms inserted within the bulk forming hydroxyl groups (O-H distance is equal to 1 {\AA}), while for the case of H insertion on the surfaces (111) and (110), there was formation of an O$-$H$\cdots$O sequence including a chemical OH bond and H-bond. Marrocchelli and Yildiz \cite{D.Marrocchelli2012} predicted that hydrogen penetration into CeO$_2$(111) was a surface-limited process with a large energy barrier (1.67 eV), while the subsequent hydrogen diffusion in the bulk ceria was rather facile with a much lower energy barrier (0.52 eV). Wu {\it et al.} \cite{X.-P.Wu2015} obtained similar results based on DFT+$U$ calculations, in the study of hydrogen diffusion at stoichiometric and reduced CeO$_2$(111) surfaces, showing that H diffusion occurred more readily at reduced CeO$_2$. Stimac and Goldman \cite{J.C.Stimac2024} performed comprehensive DFT+$U$ calculations of hydrogen diffusivity in bulk CeO$_2$. They have determined the interstitial hydrogen formation energies for three different sites, {\it i.e.}, the interstitial ({\it int}) site consisting of a hydroxyl (OH) species with a bond length of $\sim$1.0 {\AA}, the octahedral ({\it oct}) interstitial sites, and the linear site where the {\it int} site O-H bond being in line with the neighboring oxygen (O$-$H$\cdots$O). The diffusion hops between hydrogen interstitial sites have been determined, and the activation energy barriers are uniformly low (on the order of 0.1 eV or less), suggesting that hydrogen is highly diffusive in CeO$_2$.

The experimental data of the hydrogen penetration and diffusion behaviour in bulk ceria has been very limited. Schweke {\it et al.} \cite{D.Schweke2020} studied the effect of temperature, H$_2$ partial pressure and the presence of oxygen vacancies on CeO$_2$-H$_2$ interaction, showing that the penetration of hydrogen into the lattice occurred preferentially at low temperatures. Very recently, Mao {\it et al.} \cite{W.Mao2024} studied hydrogen diffusion in ceria thin films, by quantitatively determining the H depth distribution with resonant nuclear reaction analysis (NRA). The H diffusion coefficient in the ceria films was determined as $>$10$^{-18}$ m$^2$s$^{-1}$ at 773$\sim$973 K, with the activation energy of $<$1.69 eV larger than previous DFT calculations.

Both the solid state NMR and quasi-elastic neutron scattering are powerful tools for the study of diffusion dynamics of hydrogen. An advantage of the latter is the ability to detect the spatial information or the geometry of the diffusion process. In this work, a combination of NMR, X-ray total scattering and PDF analysis, neutron powder diffraction, QENS and theoretical calculations have been performed, in order to exclusively study the hydrogen mobility in bulk ceria.

\section{\uppercase\expandafter{\romannumeral2}. Methodology}

Ceria powder (of purity 99.95\%) was commercially purchased from Sigma-Aldrich. Further characterization using neutron powder diffraction at room temperature was performed using the high-resolution powder diffractometer Xuanwu \cite{Y.Xia2022} at China Mianyang Research Reactor (CMRR), with a neutron wavelength of 1.8846 {\AA} generated from Ge (115) single crystal. Rietveld refinements were performed by the program suite FULLPROF. The lattice parameter was determined as $a_0=5.413$ {\AA}. For the study of the CeO$_2$-H system, the CeO$_2$ powder sample was first activated at 673 K under vacuum for 1 h, and further exposed to 1 bar of H$_2$ at different temperatures $T=$533, 623, and 673 K for 2 h, respectively, which was analogous to that described in the previous study \cite{Z.Wu2017}.

\textbf{$^1$H $T_2$ Relaxation}

The activation energies and protonic diffusion coefficients of the CeO$_2$-H systems were determined by $^1$H NMR technique. Transverse relaxation time $T_2$ measurements were performed by using a Niumai NMR spectrometer (VTMR20-010V-I) equipped with a NMR tube with diameter $\phi$=10 mm. The frequency of the spectroscopy is 21.96 MHz and measurements were conducted under temperature from 203 K to 333 K. The temperature of the sample was stabilized by a gas-flow cryostat with liquid nitrogen and the accuracy was $\pm$0.3 K. Other key parameters include sampling frequency of 5000 KHz, 90$^\circ$ pulse width of 2.60 $\mu$s, 180$^\circ$ pulse width of 4.40 $\mu$s, radio frequency (RF) delay of 0.003 ms, and the waiting time of 3000 ms. The pulse sequence applied in this work for the measurements of $T_2$ is MSE-CPMG sequence \cite{S.Meiboom1958} as shown in Figure 1.

\begin{figure}
\centering
\includegraphics[width=8.6cm]{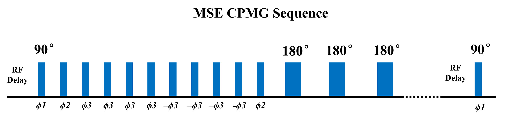}
\hspace{0.5cm}
\caption{(Color online) The MSE-CPMG sequence used in transverse relaxation time measurements.}
\end{figure}
\textbf{Quasi-elastic neutron scattering}

The large incoherent neutron scattering cross section of hydrogen makes QENS an ideal tool for probing the diffusion of hydrogen atoms in the lattice. For the ceria sample exposed to H$_2$ at 673 K for 2 h, the hydrogen concentration in CeO$_2$H$_y$ was volumetrically determined as $y=0.036$ by monitoring pressure changes in a calibrated volume. The diffusion dynamics of hydrogen in bulk CeO$_2$H$_{0.036}$ has been examined by quasi-elastic neutron scattering. The measurements were performed on the triple-axis spectrometer Yinglong at CMRR by fixing the final neutron energy $E_\mathrm{f}=4.5$ meV. The PG(002) monochromator and analyzer were used, with horizontal collimation of 20$^\prime$. The energy resolution full widths at half-maximum (FWHM) is $\sim$160 $\mu$eV determined by a standard vanadium sample. A momentum transfer ($Q$) range of $0.85\leq Q\leq2.2$ \AA$^{-1}$ was used for the constant-$Q$ scans.

\textbf{X-ray total scattering}

X-ray total scattering experiments of three different hydrogen-treated ceria samples have been performed at room temperature on the beamline BL13SSW at Shanghai Synchrotron Radiation Facility. The X-rays were monochromatized to give a wavelength of 0.248 {\AA}. The samples were loaded into thin-walled borosilicate capillary tube with diameter of 1.5 mm. In addition to the sample measurements, separate measurements were performed on empty capillary for background subtraction purpose. A standard sample CeO$_2$ (of purity 99.95\%) was used to calibrate the distance between the sample and the detector. Data were collected using a 2D iRay Mercu1717V detector for 400 s using the rapid acquisition PDF method (RAPDF) \cite{P.J.Chupas2003}, with raw diffraction data processed into one-dimensional diffraction patterns using the Dioptas program \cite{C.Prescher2015}. The PDFgetX3 program \cite{P.Juhas2013} was used for PDF data processing.

\textbf{Theoretical calculations}

The hydrogen behavior in the bulk phase of CeO$_2$ was investigated using Hubbard-corrected density-functional theory (DFT+$U$) as implemented in the Vienna $\textit{ab initio}$ simulation package (VASP) version 5.4.4 \cite{G.Kresse1996}. The exchange-correlation functional was treated within the generalized gradient approximation (GGA) using the Perdew-Burke-Ernzerhof (PBE) parametrization \cite{J.P.Perdew1996}, while the electron-ion interactions were described using the projector augmented wave (PAW) method \cite{P.E.Blochl1994}. The official PAW pseudopotentials are used, with the valence electronic configuration of Ce, O and H being $5s^26s^25p^65d^14f^1$, $2s^24p^4$, $1s^1$, respectively. To account for strong electron correlations in CeO$_2$, we employed Dudarev's DFT$+U$ scheme \cite{S.L.Dudarev1998} with a Hubbard parameter $U=4$ eV. Previous studies \cite{B.Ao2021} have shown that variations in $U$ within the range of 3$\sim$5 eV have negligible effects on the relative stability of hydrogen. To avoid metastable states in DFT+$U$, we applied the Random Orbital-Dependent Local Perturbation method \cite{R.Qiu2025}. For computational simplicity, all calculations were performed in a ferromagnetic configuration, as the relative energies were proved to be independent of the magnetic state. The cut-off energy of 500 eV in a plane-wave basis expansion and a $8\times 8\times8$ Monkhorst-Pack $k$ mesh for the unit cell was found to provide satisfactory convergence, and a $3\times 3\times 3$ Monkhorst-Pack $k$-mesh for the $2\times 2\times 2$ supercell. Electronic relaxation was performed until the total energy was converged to 1$\times$10$^{-8}$ eV, and the ionic relaxation was performed until the Hellmann-Feynman forces were less than 0.01 eV/{\AA}.

For the {\it ab initio} molecular dynamics (AIMD) simulations of hydrogen diffusion in CeO$_2$, preliminary equilibration runs were performed to stabilize the systems at the target temperatures and minimize stress tensor components. Simulations were conducted at 333, 433, 623, and 673 K, each spanning 50 ps with a time step of 0.5 fs. Thermodynamic equilibrium was achieved within the first 10 ps in all cases. The hydrogen diffusion coefficient ($D_\mathrm{AIMD}$) was extracted by fitting the slope of the mean square displacement (MSD) versus time ($t$) using Einstein's random-walk law $D_\mathrm{AIMD}=\lim_{t\rightarrow\infty}\frac{\langle\vert {\bf r}(t)-{\bf r}(0)\vert\rangle}{6t}$, where ${\bf r}(0)$ and ${\bf r}(t)$ denote the atomic positions at the initial time $t_0$ and time $t$, respectively. From the temperature-dependent diffusion coefficients, the activation energy barrier $E_\mathrm{a}$ for hydrogen diffusion was determined via an Arrhenius analysis, using a least-squares fitting procedure.
\section{\uppercase\expandafter{\romannumeral3}. Results and Discussions}

\subsection{A. $^1$H NMR $T_2$ measurements}

The $^1$H NMR $T_2$ for three different hydrogen-treated CeO$_2$ samples has been measured in the temperature range of 203 K$\leq T\leq$333 K. 
With the assumption that in the middle temperature region (253 K$<T<$293 K), there is an almost linear relationship between the FWHM $\Delta\nu$ and the transverse relaxation time $T_2$ \cite{H.Chen1989}, $2\pi\Delta\nu=\frac{2}{T_2}$. In the CeO$_2$-H system, as $^{58}$Ce and $^{16}$O being non-spin nuclei, and the low natural abundance of $^{17}$O of 0.037\%, one can thus only consider the homonuclear dipolar-dipolar ($^1$H-$^1$H) interaction, $1/T_2$ can be simply written as 
\begin{equation}\label {eq1}
\ln{\frac{1}{T_2}}=\ln F\tau_0+\frac{E_\mathrm{a}}{k_\mathrm{B}T},
\end{equation}
where the constant $F$ is related to $^1$H, $\tau_0$ is the pre-exponential factor of the correlation time $\tau_\mathrm{NMR}=\tau_0\exp{(\frac{E_\mathrm{a}}{k_\mathrm{B}T})}$, $E_\mathrm{a}$ is the activation energy, $k_\mathrm{B}$ is the Boltzman constant, and $T$ is the absolute temperature.
According to the temperature dependence of transverse relaxation time $T_2$, in the temperature range of 253 K$\sim$293 K, the linear relationship also exists between $\ln{\frac{1}{T_2}}$ and 1000/$T$. The activation energy $E_\mathrm{a}$ of CeO$_2$ exposed to H$_2$ at different temperatures can be estimated from the slope of the curves \cite{H.Chen1989,D.Zhou1987}, as listed in Table I. These values are in good agreement with the low activation energy (0.22 eV) reported by Chen {\it et al.} \cite{H.-T.Chen2007}, although much larger than the results ($<0.15$ eV) obtained by Stimac and Goldman \cite{J.C.Stimac2024} as well as our {\it ab initio} calculations.

\begin{table}[htbp]
\begin{threeparttable}
\caption{The activation energy $E_\mathrm{a}$ and the pre-exponential factors $\tau_0$ and $D_0$ of CeO$_2$ exposed to 1 bar H$_2$ at different temperatures (533, 623, and 673 K), determined by the $^1$H NMR transverse relaxation time $T_2$ measurements.}
\begin{tabular}{ccccc}
\toprule
                              &                              & 533 K                   &  623 K                   & 673 K  \\
\hline
$E_\mathrm{a}$ (eV)           &                              & 0.1905                  &  0.2013                  & 0.2143 \\
$\tau_0$(s)                   & $r_\mathrm{H}=2.75$ {\AA}    & 4.383$\times10^{-10}$   & 8.202$\times10^{-10}$    & 1.295$\times10^{-9}$\\
                              & $r_\mathrm{H}=3.80$ {\AA}    & 3.047$\times10^{-9}$    & 5.704$\times10^{-9}$     & 9.017$\times10^{-9}$\\
$D_0$ (m$^2/$s)               & $r_\mathrm{H}=2.75$ {\AA}    & 5.404$\times10^{-11}$   & 2.888$\times10^{-11}$    & 1.829$\times10^{-11}$\\
                              & $r_\mathrm{H}=3.80$ {\AA}    & 7.773$\times10^{-12}$   & 4.153$\times10^{-12}$    & 2.627$\times10^{-12}$\\
\toprule\\
\end{tabular}
\end{threeparttable}
\end{table}

Once the activation energy $E_\mathrm{a}$ has been obtained, one can further derive the correlation time $\tau_\mathrm{NMR}$ of $^1$H by using the well known Bloembergen-Purcell-Pound (BPP) formula \cite{A.Abragam1961,D.Besghini2019}
\begin{equation}\label {eq2}
\frac{1}{T_2}=\frac{3}{320}\frac{\mu_0^2\gamma_\mathrm{H}^4\hbar^2}{\pi^2r_\mathrm{H}^6}\tau_\mathrm{NMR}\bigg\{3+\frac{5}{1+\omega_0^2\tau_\mathrm{NMR}^2}+\frac{2}{1+4\omega_0^2\tau_\mathrm{NMR}^2}\bigg\},
\end{equation}
where $\mu_0=4\pi\times10^{-7}$ H$/$m is the permeability of free space, $\gamma_\mathrm{H}=2.675\times10^8$ rad$/$s$/$T is the gyromagnetic ratio, $\hbar=h/2\pi$ and $h=6.626\times10^{-34}$ J$\cdot$s is the Planck constant, $\omega_0$ is the Larmor frequency associated with the precession of the spin caused by the magnetic field, and $r_\mathrm{H}$ denotes the distance between the neighboring H nuclei. The homonuclear dipolar-dipolar interaction decreases with the increase of $r_\mathrm{H}$, so do the diffusion coefficients. However, because of lack of detailed structural information of proton positions in the crystal structure of CeO$_2$-H systems, two selected values of $r_\mathrm{H}$ have been assumed according to DFT calculations \cite{X.-P.Wu2015}, {\it i.e.}, at CeO$_2$(111) the distance between two neighboring top-surface O atoms of $\sim3.8$ {\AA}, and that between neighboring top- and sub-surface O atoms of $\sim2.75$ {\AA}. Fitting to the temperature dependence of the transverse relaxation time $T_2$ was performed by using the BPP equation \ref{eq2}, as shown in Figure 2, with the fitted pre-exponential factor $\tau_0$ listed in Table I.
\begin{figure}
\centering
\includegraphics[width=8.6cm]{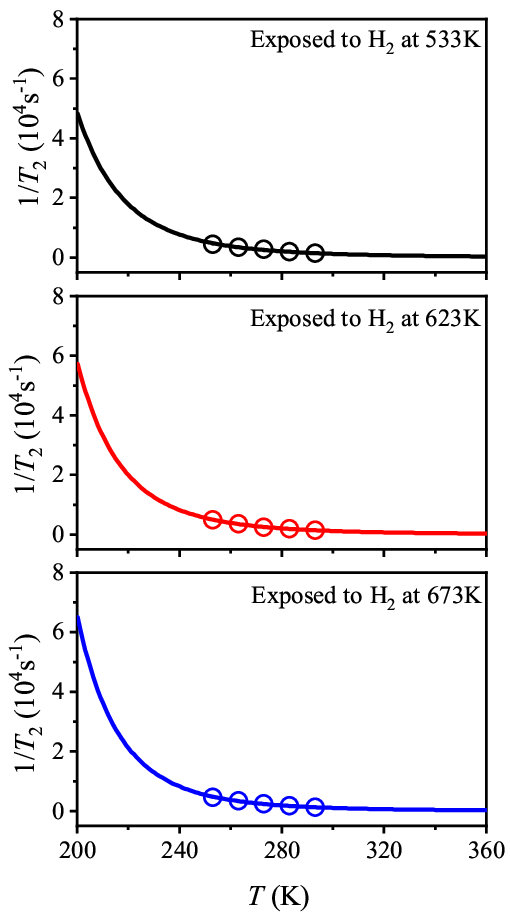}
\hspace{0.5cm}
\caption{(Color online) Temperature dependence of $^1$H NMR $T_2$ for three ceria samples exposed to H$_2$ at (a) $T=533$ K, (b) $T=623$ K, (c) $T=673$ K. A fitting of BPP type formula, Eq.2 is shown as a solid line.}
\end{figure}

Considering the random Brownian motion, the diffusion constant obtained by NMR, $D_\mathrm{NMR}$ can be related to $\tau_\mathrm{NMR}$ by the three-dimensional Einstein-Smoluchowski equation $D_\mathrm{NMR}=\frac{l_\mathrm{NMR}^2}{6\tau_\mathrm{NMR}}$, where $l_\mathrm{NMR}$ denotes the jump distance of protons. Based on DFT calculations, forming OH group is the most stable existence state for hydrogen in ceria, and the hopping between the neighbor oxygen sites can readily occur, thus $l_\mathrm{NMR}=3.8$ {\AA} has been estimated.

Figure 3 shows the derived H diffusion coefficients $D_\mathrm{NMR}=D_0\exp{(\frac{E_\mathrm{a}}{k_\mathrm{B}T})}$ in the three hydrogen-treated ceria samples as a function of temperature, with the pre-exponential factor $D_0$ listed in Table I. In the temperature range $T>300$ K, the H diffusion coefficients are found to depend on the hydrogen-treated conditions, {\it i.e.}, the ceria sample exposed to H$_2$ at the highest temperature 673 K has the largest $D_\mathrm{NMR}$, while the sample exposed to H$_2$ at the lowest temperature 533 K has the lowest diffusion coefficients. This tendency could be related to different hydrogen concentrations in these three samples.
\begin{figure}
\centering
\includegraphics[width=8.6cm]{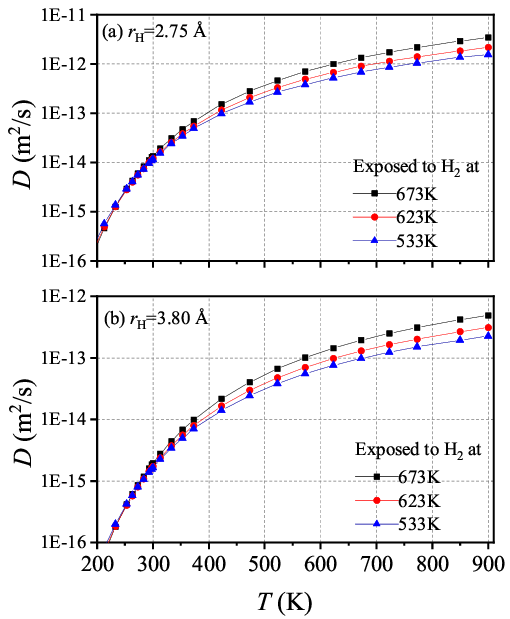}
\hspace{0.5cm}
\caption{(Color online) Plots of the H diffusion coefficients derived from $^1$H NMR $T_2$ measurements in three different hydrogen-treated CeO$_2$ samples as a function of temperature from 200 K to 900 K, with the distance between the neighboring H nuclei assumed as (a) $r_\mathrm{H}=2.75$ {\AA} and (b) $r_\mathrm{H}=3.8$ {\AA}. }
\end{figure}

\subsection{B. Neutron powder diffraction}

Neutron powder diffraction measurements have been performed at room temperature for both the pure ceria and the hydrogen-treated ceria samples. As shown in the diffraction patterns in Figure 4, new well-defined peaks appear in the vicinity of the characteristic peak (400) of ceria, indicating the presence of new sub-phases. Based on Rietveld refinements, the main phase CeO$_2$ ($a_0=5.413$ {\AA}) do not exhibit oxygen vacancies. These two sub-phases have been assigned to the cubic oxyhydroxide structure of CeO$_2$H$_y$ ($0\le y\le1$, $a_0=5.440$ {\AA}), and the non-stoichiometric composition of CeO$_{1.34}$ ($a_0=5.461$ {\AA}). These results show good consistency with the previous neutron powder diffraction experiments \cite{T.Matsukawa2021}, where two new peaks close to the Bragg reflection (220) of ceria have been observed and determined to be the (220) reflections of sub-phases CeO$_2$H and CeO$_{1.84}$. In this work, the contributions from the two sub-phases can be clearly distinguished from the (400) peak, while there are no obvious additional peaks close to (220) peak of CeO$_2$ in the pattern. The mass fractions of the main phase CeO$_2$, secondary phase CeO$_2$H, and third phase CeO$_{1.34}$ are estimated as 97.37\%, 0.59\%, 2.04\%, respectively.

\begin{figure}
\centering
\includegraphics[width=8.6cm]{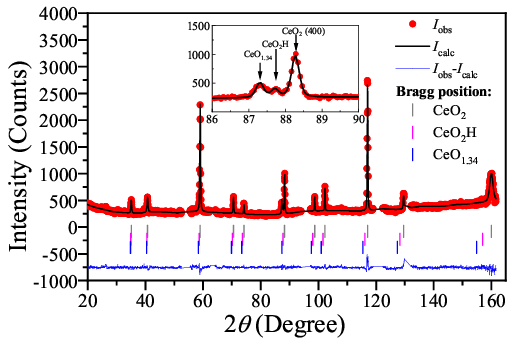}
\hspace{0.5cm}
\caption{(Color online) Neutron powder diffraction patterns of the hydrogen-treated ceria sample. The red solid points, black solid curve, blue solid curve, grey ticks, pink ticks, and blue ticks indicate experimental data $I_\mathrm{obs}$, calculated data $I_\mathrm{calc}$, the difference between the experimental and calculated intensities $I_\mathrm{obs}$-$I_\mathrm{calc}$, the positions of allowed reflections in CeO$_2$, CeO$_2$H, and CeO$_{1.34}$, respectively. The inset figure displays the fitting pattern around the (400) peaks.}
\end{figure}
\subsection{C. Quasi-elastic neutron scattering}
Since the incoherent scattering cross-section of hydrogen [$\sigma_\mathrm{inc}$(H)=80.26 barn] is very large, compared with neglectable values of both cerium $\sigma_\mathrm{inc}$(Ce) and oxygen $\sigma_\mathrm{inc}$(O), the observed neutron scattering from CeO$_2$H$_{0.036}$ is almost exclusively due to the hydrogen. As an example of the data, Fig. 5(a) shows the dynamical structure factor recorded at $Q=1.2$ {\AA}$^{-1}$ and $Q=1.8$ {\AA}$^{-1}$. The incoherent QENS spectra $S(Q,\omega)$ were fitted to the following function \cite{M.Kofu2020},
\begin{equation}\label {eq3}
S(Q,\omega)=R(Q,\omega)\otimes\bigg[\frac{e^{-\frac{1}{3}\langle u^2\rangle Q^2}}{\pi}\frac{\Gamma(Q)}{\omega^2+\Gamma(Q)^2}\bigg],
\end{equation}
where $R$($Q,\omega$) is the resolution function of the instrument, the prefactor $e^{-\frac{1}{3}\langle u^2\rangle Q^2}$ is the Debye-Waller factor where $\langle u^2\rangle$ is the mean-square displacement of protons, and $\Gamma$($Q$) is the FWHM of a Lorentzian QENS function centered at the energy transfer $\hbar\omega=0$. The $Q$ dependence of the width of the quasi-elastic component is shown in Fig. 5(b), and the diffusion process can be well described by the well-known Chudley and Elliott (CE) model \cite{C.T.Chudley1961},
\begin{equation}\label {eq4}
\Gamma(Q)=\frac{\hbar}{\tau}\bigg(1-\frac{\sin Ql}{Ql}\bigg)=\frac{6\hbar D_\mathrm{s}}{l^2}\bigg(1-\frac{\sin Ql}{Ql}\bigg),
\end{equation}
where $\tau$ represents the mean residence time the hydrogen atom spends at a site before it jumps again to another site, $l$ is the characteristic jump distance, and $l^2/6\tau=D_\mathrm{E}$ is the so-called Einstein diffusion coefficient; however, at finite concentration it has to be substituted by the self-diffusion coefficient $D_\mathrm{s}$ \cite{R.Hempelmann}. In the CE model, it is assumed that the diffusion process consists of a sequence of elementary jumps of atoms into adjacent vacant sites, and the jumps are instantaneous ({\it i.e.}, jumping in a time $t\ll\tau$) and uncorrelated with the vibrational motions. By fitting with eq.\ref{eq4}, one can obtain the self-diffusion coefficient of hydrogen $D_\mathrm{s}=(2.7\pm0.8)\times10^{-9}$m$^2/$s, and the jump length $l=(3.98\pm0.56)$ {\AA} which just corresponds to the distance between the next nearest neighboring oxygen sites. The results suggest that the hydrogen diffusion mechanism may consist of hydrogen reorientation of the hydroxyl (OH$\rightarrow$O$-$H$\cdots$O), hydrogen jumping to the adjacent oxygen (O$-$H$\cdots$O$\rightarrow$OH), and subsequent reorientation of the hydroxyl, which shows good agreements with our theoretical calculations.
\begin{figure*}
\centering
\includegraphics[width=17.4cm]{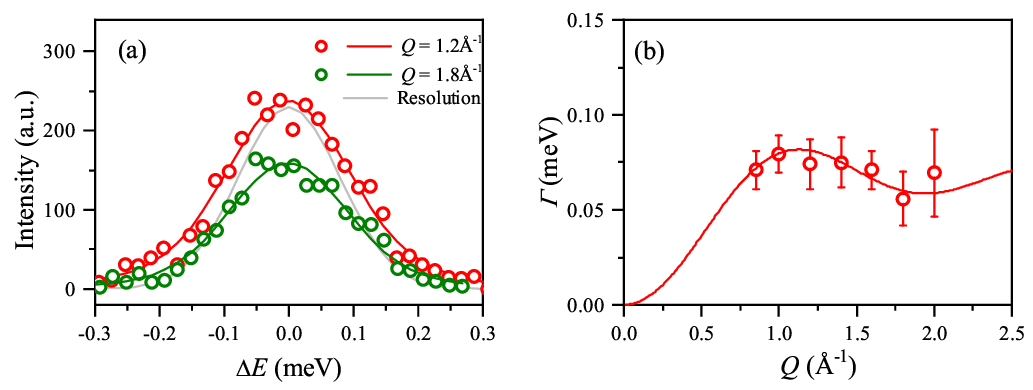}
\hspace{0.5cm}
\caption{(Color online) (a) The dynamic structure factor of CeO$_2$H$_{0.036}$ observed at different $Q$. (b) The $Q$ dependence of HWHM $\Gamma$ in CeO$_2$H$_{0.036}$, the solid curve is the results of the fit based on the CE model. }
\end{figure*}
In the low-$Q$ region, $\Gamma(Q)=\hbar D_\mathrm{s}Q^2$ is generally valid irrespective of the details of the diffusion process, {\it i.e.}, the diffusion obeys Fick's law, where $D_\mathrm{s}=l^2/6\tau$ is the long-range self-diffusion coefficient. However, it should be noted that there exists very large discrepancy between the diffusion coefficients obtained by QENS and the $^1$H NMR $T_2$ measurements, which may suggest that different techniques can observe protonic dynamics having different characteristic times \cite{A.Ishikawa2008} within the complicated CeO$_2$-H interacting system.

\subsection{D. X-ray total scattering}
To look for possible evidence of the local structure of the CeO$_2$-H systems, synchrotron X-ray total scattering experiments have been performed. The atomic pair distribution function (PDF), denoted as $G(r)$, is obtained by a Fourier transformation of the powder diffraction data according to eq.\ref{eq5},
\begin{equation} \label {eq5}
G(r)=\frac{2}{\pi}\int^{Q_\mathrm{max}}_{Q_\mathrm{min}}{Q\big[S(Q)-1\big]\sin(Qr)\mathrm{d}Q},
\end{equation}
where $r$ denotes the distance between two atoms in real space, $Q$ is the magnitude of the momentum transfer, and $S(Q)$ is the total scattering structure function \cite{T.Egami2003}. Because of the unfavorable signal to noise ratio in the high-$Q$ region, $Q[S(Q)-1]$ has been truncated at $Q_{\mathrm{max}}=21$ {\AA}$^{-1}$ before the transformation. The real space modeling is carried out using the program PDFgui \cite{C.L.Farrow2007} for the local structural studies. Figure 6 shows the X-ray PDFs and the fitted results using the crystallographically constrained fluorite structure model, which does not include H atoms as the signal can be negligible. The scale factor, lattice parameters, isotropic atomic displacement parameters ($U_{\mathrm{iso}}$) for Ce and O atoms, and the peak sharpening parameter have been refined. The $Q_{\mathrm{broad}}$ and $Q_{\mathrm{damp}}$ parameters characterizing the broadening and the exponential decay of PDF peaks due to instrumental resolution are fixed at values obtained by the standard sample.

\begin{figure*}
\centering
\includegraphics[width=17.4cm]{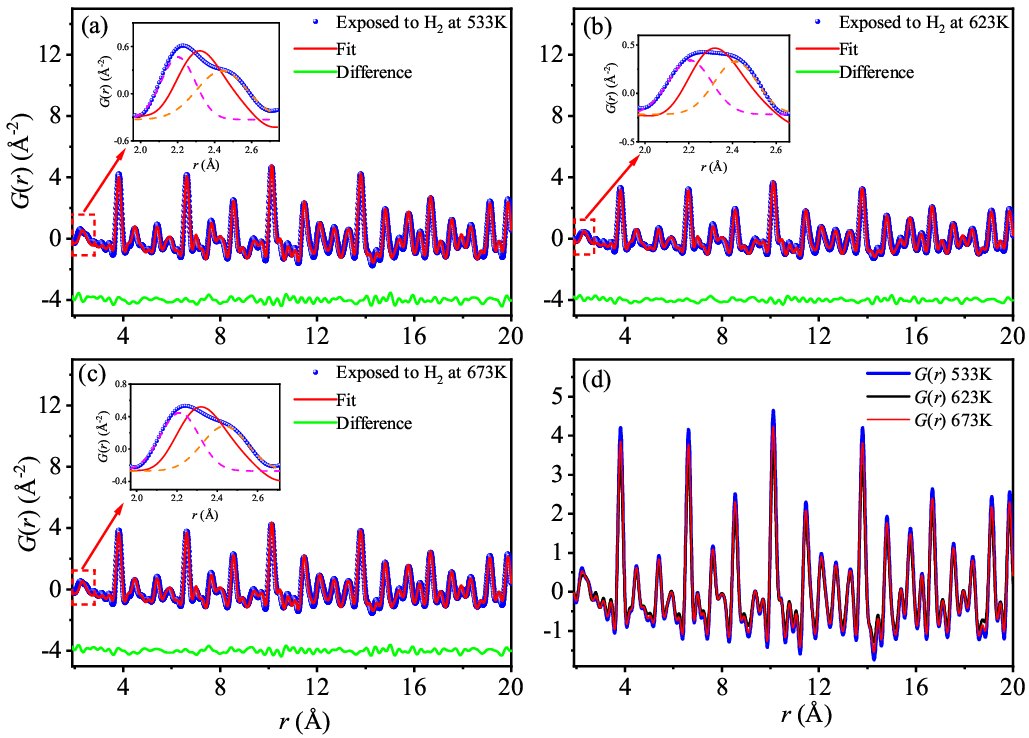}
\hspace{0.5cm}
\caption{(Color online) The PDF for three CeO$_2$ samples exposed to H$_2$ at (a) $T=533$ K, (b) $T=623$ K, (c) $T=673$ K, measured by synchrotron X-ray total scattering (blue circles) and fitted by the averaged structure (red lines). The insets display the double-peak characters centered at $\sim$2.3 {\AA} corresponding to the Ce-O correlation, which cannot be fitted by the averaged structure. (d) A direct comparison of the PDF for the three different hydrogen-treated CeO$_2$ samples.}
\end{figure*}

It can be seen from Figure 6(a)-(c) that the $G(r)$ curves of the three CeO$_2$-H samples can be well fitted by the average structure model, and they look very similar as shown in Figure 6(d). From Rietveld refinement, the lattice parameter is around $a=5.41$ {\AA}, consistent with the results of neutron diffraction measurements. 
The first and second PDF peaks centered at $\sim$2.3 {\AA} and $\sim$3.8 {\AA} correspond to Ce-O and Ce-Ce atomic pairs, respectively. Overall, the average structure model fits the short-medium range very well for both the position and intensity of the peaks in $G(r)$ curves, with no clear mismatch in the region of $r<20$ {\AA} for the three samples. However, it should be noted that there is one exception, {\it i.e.}, the first peak in $G(r)$ curves corresponding to the Ce-O correlation exhibits a double-peak feature, as shown in the insets of Figure 6. Taking the hydrogen-treated sample at 673 K as an example, the intensity contributed by Ce-O correlation cannot be described by the average structure model, but well fitted by two Gaussian functions centered at 2.21 {\AA} and 2.44 {\AA}, respectively. The deviations from the average structure model could be due to the lattice distortion deviated from the equilibrium state induced by hydrogen incorporation and oxygen vacancies. 

\subsection{E. {\it Ab initio} calculations}
CeO$_2$ crystallizes in the $Fm\bar{3}m$ fluorite structure (space group No.225). The optimized lattice parameter is obtained as $a_0=5.48$ {\AA}, and an equilibrium atomic volume of $V_0=164.90$ {\AA}$^3$, as listed in Table II, are in good agreement with the experimental and theoretical results. In order to determine the favorable existence states of hydrogen, the hydrogen formation energies $E_\mathrm{F}$ in bulk CeO$_2$ for the octahedral interstitial site (denoted as H$_{\mathrm{Int}}$), hydroxyl group (H$_{\mathrm{OH}}$), as well as the hydrogen atom occupying the oxygen vacancy (H$_{\mathrm{OV}}$) have been evaluated, respectively. The standard definition of $E_\mathrm{F}$ can be written as \cite{B.Ao2021}
\begin{eqnarray}
E_\mathrm{F}(\mathrm{H}_{\mathrm{Int}})&=&E_{\mathrm{Tot}}(\mathrm{Ce}_x\mathrm{O}_y\mathrm{H}) \nonumber\\
&-&E_\mathrm{Tot}(\mathrm{Ce}_x\mathrm{O}_y)-\frac{1}{2}E_{\mathrm{Tot}}(\mathrm{H}_2),
\end{eqnarray}
\begin{eqnarray}
E_\mathrm{F}(\mathrm{H}_{\mathrm{OH}})&=&E_{\mathrm{Tot}}(\mathrm{Ce}_x\mathrm{O}_y\mathrm{H})  \nonumber\\
&-&E_{\mathrm{Tot}}(\mathrm{Ce}_x\mathrm{O}_y)-\frac{1}{2}E_{\mathrm{Tot}}(\mathrm{H}_2),
\end{eqnarray}
\begin{eqnarray}
E_\mathrm{F}(\mathrm{H}_{\mathrm{OV}})&=&E_{\mathrm{Tot}}(\mathrm{Ce}_x\mathrm{O}_{y-1}\mathrm{H})+\frac{1}{2}E_{\mathrm{Tot}}(\mathrm{O}_2) \nonumber\\
&-&E_{\mathrm{Tot}}(\mathrm{Ce}_x\mathrm{O}_y)-\frac{1}{2}E_{\mathrm{Tot}}(\mathrm{H}_2).
\end{eqnarray}

\begin{table}[htbp]
\begin{threeparttable}
\caption{The optimized lattice parameter $a_0$, and unit cell volume of CeO$_2$, compared with experimental and DFT calculations.}
\begin{tabular}{ccc}
\toprule
                                      & $a_0$ (\AA)     &  $V_0$ (\AA$^3$)\\
\hline
Expt.\cite{L.Gerward2005}             & 5.41            &  158.43\\
GGA \cite{M.Nakayama2009}             & 5.46            &  163.04\\
GGA+$U$ \cite{P.R.L.Keating2012}      & 5.49            &  165.83\\
GGA+$U$ \cite{J.L.F.Da Silva2007}     & 5.49            &  165.47\\
DFT+$U$ (this work)                   & 5.48            &  164.90\\
Neutron diffraction (this work)       & 5.41            &  158.58\\
X-ray total scattering (this work)    & 5.41            &  158.08\\
\toprule\\
\end{tabular}
\end{threeparttable}
\end{table}

The total energies and the formation energies of the CeO$_2$-H system are listed in Table III, which are in good agreement with the previous study \cite{B.Ao2021}. It can be shown that $E_\mathrm{Tot}$(H$_\mathrm{OH}$)$<E_\mathrm{Tot}$(H$_\mathrm{Int}$)$<$$E_\mathrm{Tot}$(CeO$_2$)$<E_\mathrm{Tot}$(H$_\mathrm{OV}$), and $E_\mathrm{F}$(H$_\mathrm{OH}$)$<E_\mathrm{F}$(H$_\mathrm{Int}$)$<E_\mathrm{F}$(H$_\mathrm{OV}$), with the hydroxyls being energetically most stable. The climbing image nudged elastic band (CI-NEB) method \cite{G.Henkelman2000} has been employed to search for the minimum energy pathways (MEP) connecting two local minima. Figure 7 presents the results of CI-NEB calculations for the hydrogen atom rotating around one oxygen atom from OH to O$-$H$\cdots$O existing state, and hydrogen atom with O$-$H$\cdots$O configuration moving to the nearest neighbour oxygen atom with the OH, plotted by VESTA software \cite{K.Momma2011}. Figure 8 shows the calculated energy profiles for H diffusions. The corresponding energy barriers, or the energy differences between the transition state (TS) and the O$-$H$\cdots$O existing state are relatively small as $\sim$0.07 eV and $\sim$0.08 eV, respectively. The result of the hydrogen rotation is comparable to a very recent study \cite{J.C.Stimac2024}, where the activation energy has been calculated from PBE, PBEsol, and SCAN exchange correlation functionals ranging from 0.03$\sim$0.06 eV. It is shown that hydrogen can readily diffuse through the CeO$_2$ lattice between neighbour oxygen sites.

\begin{figure*}
\centering
\includegraphics[width=17.4cm]{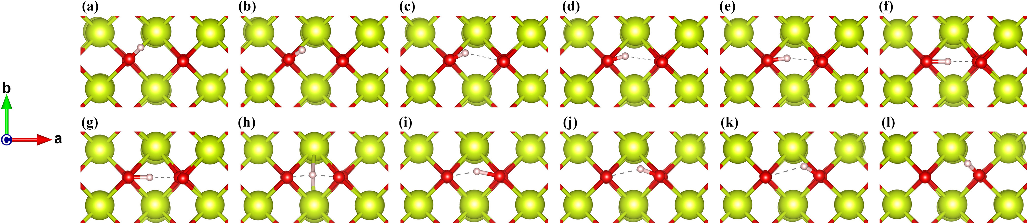}
\hspace{0.5cm}
\caption{(Color online) Critical point configurations for OH to O$-$H$\cdots$O (top row) and O$-$H$\cdots$O to OH (bottow row) MEPSs. }
\end{figure*}

\begin{table*}[htbp]
\begin{threeparttable}
\caption{The total energies $E_{\mathrm{Tot}}$ (eV) and formation energies $E_\mathrm{F}$ (eV) of the CeO$_2$-H system, compared with previous DFT calculations \cite{B.Ao2021}.}
\begin{tabular}{cccccccc}
\toprule
                       & $E_\mathrm{Tot}$(CeO$_2$)     & $E_\mathrm{Tot}$(H$_{\mathrm{Int}}$) & $E_{\mathrm{Tot}}$(H$_{\mathrm{OH}}$) & $E_{\mathrm{Tot}}$(H$_{\mathrm{OV}}$)   & $E_\mathrm{F}$(H$_{\mathrm{Int}}$)    & $E_\mathrm{F}$(H$_\mathrm{OH}$)     & $E_\mathrm{F}$(H$_{\mathrm{OV}}$)\\
\hline
Ref. \cite{B.Ao2021}   & -788.84                & -789.80              & -791.49             &   $-$                   & 2.41                & 0.72                & 3.92\\
This work              & -792.36                & -793.02              & -794.68             & -787.20               & 2.73                & 1.06                & 3.26\\
\toprule\\
\end{tabular}
\end{threeparttable}
\end{table*}

\begin{figure}
\centering
\includegraphics[width=8.6cm]{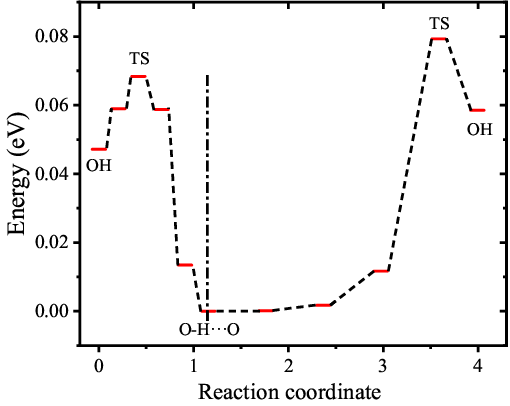}
\hspace{0.5cm}
\caption{(Color online) Minimum energy path for H diffusion through interacting with neighboring oxygen atoms, predicted by the CI-NEB calculations. The relevant reaction configurations are shown in Figure 7.}
\end{figure}

According to {\it ab initio} MD simulations, in the temperature range of 333$\sim$673 K, the hydrogen diffusion in CeO$_2$-H system follows an Arrhenius-type relation $D=D_0\exp(-\frac{E_\mathrm{a}}{k_\mathrm{B}T})$. In Figure 9, the diffusivity data are presented as a function of temperature, and the fitting parameters have been obtained from the plot. The activation energy $E_\mathrm{a}=0.044$ eV which shows good consistency with our CI-NEB calculations, further supporting that relatively low energy is required for hydrogen to overcome the jump barrier, thus the migration of H atoms is accessible in the bulk ceria. The prefactor $D_0=2.9\times10^{-9}$ m$^2/$s, and the diffusion coefficient at room temperature agrees well with that obtained from QENS measurements, {\it i.e.}, in order of magnitude of 10$^{-9}$ m$^2/$s.
\begin{figure}
\centering
\includegraphics[width=8.6cm]{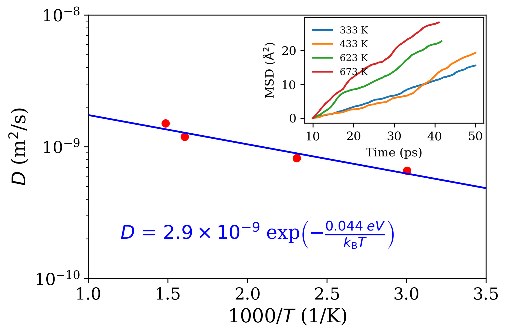}
\hspace{0.5cm}
\caption{(Color online) Hydrogen diffusivity values (red solid cycles) for CeO$_2$-H presented as a function of temperature, the solid line is the Arrhenius fit to the data. The inset shows the AIMD simulated mean square displacement (MSD) versus time at different temperatures.}
\end{figure}

\section{\uppercase\expandafter{\romannumeral4}. CONCLUSIONS}

The hydrogen diffusion dynamics in ceria has been investigated by means of $^1$H NMR, X-ray$/$neutron scattering and {\it ab initio} calculations. The QENS measurements indicate that hydrogen dynamics in CeO$_2$H$_{0.036}$ can be well described by the Chudley-Elliott jump model. The diffusion coefficient at room temperature is $D_\mathrm{QENS}=(2.7\pm0.8)\times10^{-9}$m$^2/$s, and a fixed jump length of the proton hopping motion has been obtained as $l=(3.98\pm0.56)$ {\AA}, which is very close to the distance between two next-nearest neighbor oxygen atoms $d_\mathrm{OO}=3.8$ {\AA}. The {\it ab initio} calculations also support such jump diffusions since that hydrogen can readily move in the host lattice through interacting with oxygen atoms. The estimated energy barriers $E_\mathrm{a}$ for the hydrogen atom rotating around one oxygen atom from OH to O$-$H$\cdots$O and from O$-$H$\cdots$O moving to the adjacent oxygen atom with the OH configuration are relatively small ($<0.1$ eV).

On the other hand, the $^1$H NMR transverse relaxation time $T_2$ measurements give a higher $E_\mathrm{a}$ value around 0.2 eV, and the H diffusion coefficient several orders of magnitude lower than the QENS results, suggesting some different kind of slower dynamics of protons which requires further investigations. Although X-ray is not a sensitive probe for hydrogen, and the atomic pair distribution functions $G(r)$ of the three hydrogen-treated samples look very similar, one can still find evidence of lattice distortion, from the double-peak features located at $\sim$2.3 {\AA} contributed from Ce-O correlation which cannot be fitted by the averaged structure. This work may shed light on deep understanding of the complicated CeO$_2$-H$_2$ interaction.
\section*{ACKNOWLEDGMENTS}

This work was supported by National Natural Science Foundation of China, No.12475305, No.U2430209, No.U2330104, and No.52302193. Z. Ma was supported by National Key Laboratory of Neutron Science and Technology, No.NST20240107. Z. Zhu was supported by the External Research Collaboration Project of China Petroleum and Chemical Corporation, No.U20B6003. R. Qiu was supported by the Foundation of President of China Academy of Engineering Physics, No.YZJJZQ2022011. The authors thank the BL13SSW beamline at the Shanghai Synchrotron Radiation Facility for the PDF experiments supports. The computational resources utilized in this research were provided by Hefei Advanced Computing Center, and Supercomputer Center of Institute of Computer Application, China Academy of Engineering Physics (CAEP).

\section*{Conflict of Interest}
The authors declare no conflict of interest.

\end{document}